# On-chip all-dielectric fabrication-tolerant zero-index metamaterials

**Shota Kita,[1] Yang Li,[1] Philip Muñoz,[1] Orad Reshef,[1] Daryl I. Vulis,[1] Robert W. Day, [2] Eric Mazur,[1,3] and Marko Lončar[1,*]**

1 *John A. Paulson School of Engineering and Applied Sciences, Harvard University, 9 Oxford St., Cambridge, 02138 MA*
2 *Department of Chemistry and Chemical Biology, Harvard University, 12 Oxford St., Cambridge, 02138 MA*
3 *Department of Physics, Harvard University, 29 Oxford St., Cambridge, 02138 MA*
*Corresponding author: loncar@seas.harvard.edu

**Zero-index metamaterials (ZIMs) offer unprecedented ways to manipulate the flow of light, and are of interest for wide range of applications including optical cloaking, super-coupling, and unconventional phase-matching properties in nonlinear optics. Impedance-matched ZIMs can be obtained through a photonic Dirac-cone (PDC) dispersion induced by an accidental degeneracy of two linear bands — typically an electric monopole mode and a transverse magnetic dipole mode — at the center of the Brillouin zone. Consequently, PDC can only be achieved for a particular combination of geometric parameters of the metamaterial, and hence is sensitive to fabrication imperfections. These fabrication imperfections may limit the usefulness in practical applications. In this work we overcome this obstacle and demonstrate robust all-dielectric (AD) ZIM that supports PDC dispersion over wide parameter space. Our structure, consisting of an array of Si pillars on silica substrate, is fabricated in silicon-on-insulator (SOI) platform and operates at telecom wavelengths.**

## 1. INTRODUCTION

Optical metamaterials have attracted attention because their optical properties can be engineered. In the extreme case, the refractive index can even be driven to zero [1]. The effective wavelength and phase velocity of such a zero-index metamaterial (ZIM) are infinitely large, which can lead to a plethora of interesting optical phenomena and applications. These include optical cloaking [1, 2], super coupling [3], unconventional phase-matching properties in nonlinear optics [6, 6], large mode volume single mode surface emitting lasers [8, 9], and control of the radiation characteristics of emitters [10–14]. Because the refractive index $n_\text{eff} = (\mu_\text{eff} \cdot \varepsilon_\text{eff})^{1/2}$ arises due to the combined electric and magnetic properties of a material, $n_\text{eff} = 0$ can be achieved when either $\mu_\text{eff} = 0$ (Mu-near-zero material) or $\varepsilon_\text{eff} = 0$ (Epsilon-near-zero material) [15]. However, this can cause the effective wave impedance $Z_\text{eff} (= (\mu_\text{eff}/\varepsilon_\text{eff})^{1/2})$ of the metamaterial to be either zero or infinity. In both cases the reflection of light at the metamaterial interface is large, and efficient coupling of light into the material is not possible [15, 16]. In the case when both $\mu_\text{eff} = 0$ and $\varepsilon_\text{eff} = 0$ [1], it is possible to achieve simultaneously $n_\text{eff} = 0$ and a finite $Z_\text{eff}$, reducing the reflection at the interface. One possible implementation of a simultaneously μ and ε-near-zero (MENZ) metamaterial, shown in Figure 1(a), is based on a square array of dielectric pillars. This structure can support electric monopole and transverse magnetic dipole modes that feature electric and magnetic flux loops circling around each pillar, respectively (Fig. 1(b)). Importantly, the fields from neighboring pillars cancel one another because their fields are looping in opposite directions. As a result, ***D*** and ***B*** field components vanish, which leads to effective $\mu_\text{eff} = 0$ and $\varepsilon_\text{eff} = 0$. In the structure shown in Fig. 1, this condition can be met by careful tuning of the pitch (*a*) and pillar diameter (2*r*) of the array. As a result, we can obtain a PDC as shown in Fig. 1(c). Despite several experimental demonstrations of such metamaterials [2, 10, 18] fabrication of these structures is challenging because the $\mu_\text{eff} = \varepsilon_\text{eff} = 0$ degeneracy can be achieved only for a narrow range of parameters. This limitation imposes stringent constraints on the fabrication tolerances: typically less than ±0.6% discrepancy between designed and fabricated structure is required [1].

In this work, we propose and demonstrate an all-dielectric zero-index metamaterial (AD-ZIM) in the telecom regime that is robust to fabrication imperfections. Fig. 1(a) illustrates the schematic of our AD-ZIM consisting of a square lattice of Si pillars on SiO$_2$/Si substrate. Because the pillar spacing *a* and pillar height $h_\text{Si}$ are defined by high-resolution e-beam lithography and material growth, respectively, they can be well controlled during the fabrication process (to within 5 nm).

On the other hand, the pillar diameter $2r$ can vary significantly because it is affected by processing conditions that include exposure dose, development time, and etching. Designs that are insensitive to variations in $2r$ are therefore crucial in order to enable realization of a robust ZIM. The paper is organized as follows: We first outline our design principles and present a robust ZIM design. Next, we summarize the fabrication procedure used to realize an on-chip waveguide-fed AD-ZIM prism. Then, we present our experimental measurements of refraction inside the prism, and use this information to evaluate the effective index. Finally we compare and discuss our experimental and theoretical results.

## 2. DESIGN OF FABRICATION-TOLERANT AD-ZIM

The two modes of interest in this work, shown in Fig. 1(b), are a "monopole mode" and "dipole mode" of the structure. The representative dispersion diagram for these two modes, shown in Fig. 1(c), features a photonic Dirac cone (PDC). Because each eigenfrequency of the modes forming a PDC is sensitive to the pillar geometry, including $2r$, the existence of PDC is very sensitive to fabrication imperfections [16]. For example, we find that if the fabricated $2r$ differs from the theoretical value by more than 5 nm, the degeneracy is broken, the Dirac-like dispersion is lost, and a photonic bandgap appears. This strict fabrication tolerance is a major impediment to the realization of ZIMs in the telecom and visible wavelength regime [6, 10, 18]. To overcome this impediment, we are interested in designs that are tolerant to variations in metamaterial geometry. To achieve such tolerance, both the monopole and dipole modes shown in Fig. 1(b) need to have the same eigenfrequency at the Γ-point, as well as the same dependence on structural variation (slope of $\Delta\lambda/\Delta 2r$). This sensitivity is inversely proportional to the overlap of the mode with the pillar $\eta_{Si}$. If the size of the pillar is smaller than the wavelength, the modal equivalent index $n_{eq}$ can be approximated as

$$n_{eq} \sim \eta_{Si}n_{Si} + \eta_{SiO2}n_{SiO2} + \eta_{air}n_{air}, \qquad (1)$$

where $\eta_{SiO2}$ and $\eta_{air}$ describe the overlap of the mode with $SiO_2$ and air region, respectively ($\eta_{Si} + \eta_{SiO2} + \eta_{air} = 1$). Therefore, in order to obtain an insensitive structure it is important to achieve the same modal overlap for both modes. To verify this overlap, we calculate $n_{eq}$ for both monopole and dipole modes ($n_{eq\_mono}$ and $n_{eq\_di}$) for different $h_{Si}$, and summarize the results in Fig. 2(a). In our calculations, $a$ = 879 nm, $2r$ = 512 nm, $n_{Si}$ = 3.42 and $n_{SiO2}$ = 1.45. The $n_{eq}$ of both modes are the same when $h_{Si}$ = 885 nm. Using this $h_{Si}$, we calculate the resonant wavelengths $\lambda_{mono}$ and $\lambda_{di}$ as a function of $2r$ (Fig. 2(b)) and find that the two modes have nearly identical slopes $\Delta\lambda/\Delta 2r$ for wide range of $2r$, however the degeneracy is lost. From these results, we can conclude that a robust metamaterial structure needs to satisfy two important conditions: i) $\lambda_{mono} = \lambda_{di}$ and ii) $n_{eq\_mono} = n_{eq\_di}$. To accomplish this we investigate the dependence of $\lambda_{mono}$, $\lambda_{di}$, $n_{eq\_mono}$, $n_{eq\_di}$ on important lattice parameters, and approximate this dependence with the linear approximations below:

$$\begin{cases} \lambda_i \sim A_i h_{Si} + B_i a + C_i \\ n_{eq\_i} \sim D_i h_{Si} + E_i a + F_i \end{cases}, \qquad (2)$$

where $A_i = \Delta\lambda_i/\Delta h_{Si}$, $B_i = \Delta\lambda_i/\Delta a$, $D_i = \Delta n_{eq\_i}/\Delta h_{Si}$, $E_i = \Delta n_{eq\_i}/\Delta a$, and $C_i$ and $F_i$ are free parameters, and $i$ = {mono, di}. To estimate all the coefficients in Eq. 2, we simulated each structural dependency for $a$ and $h_{Si}$. By requiring $\lambda_{mono} = \lambda_{di}$ and $n_{eq\_mono} = n_{eq\_di}$, we solve Eq. 1 for $a$ and $h_{Si}$, and find the optimal geometry to be $a$ = 967 nm and $h_{Si}$ = 897 nm. With these values, we calculate the modal wavelength dependence on $2r$ as shown in Fig. 3(a). We find much more robust operation, albeit at a longer wavelength of approximately 1700 nm. However, as the Maxwell's equations are scale invariant, as the last step we tune the operation wavelength by scaling the size of the structure as shown in Fig. 3(b). Using this approach, we derive the ideal lattice parameters to be $a$ = 918 n m and pillar height $h_{Si}$ = 860 nm for our experimental demonstration, resulting in the monopole and dipole modes that are degenerate for a range of $2r$ as shown in Fig. 4(a). Importantly, both curves have similar slopes $\Delta\lambda/\Delta 2r$, indicating a shared wavelength λ at the Γ point. If we define "near-PDC" as a near-degeneracy with $|\Delta\lambda| < 1$ nm ($|\Delta\lambda/\lambda| < 0.065\%$), the tolerance of the pillar diameter is extended up to ±19 nm ($2r$ = 508 – 546 nm). This is an order of magnitude larger tolerance window than in the case of the designs with the "accidental" degeneracy which is limited to < ±2 nm. Fig. 4(b) shows the calculated band diagrams around the Γ point for five different values of $2r$, clearly illustrating the power of our concept and structural robustness of our AD-ZIM. If $2r$ becomes either too small or too large, the degeneracy is slightly broken, as shown in the leftmost ($2r = 496$ nm) and rightmost ($2r = 560$ nm) figures. Experimentally, we expect almost bandgap-free zero-index behavior even with a large variation in $2r$. The same mechanism also allows us to tune zero-index wavelength from 1560 nm to 1620 nm simply by changing $2r$.

## 3. EXPERIMENTS AND RESULTS

To verify the zero-index behaviors and the fabrication tolerance of the AD-ZIM, we fabricated a series of AD-ZIM prisms with varying $2r$, and measure the refraction of light through these prisms. To fabricate samples, we prepared a Silicon-on-insulator (SOI) wafer with $h_{Si} = 850 - 870$ nm. To obtain this specific $h_{Si}$ we used a Si regrowth technique, depositing additional silicon by chemical vapor deposition on top of a commercially available SOI wafer with an original Si device layer thickness of 510 nm. Next, we fabricate pillar arrays in the shape of an isosceles right triangle using e-beam lithography followed by an inductively coupled plasma reactive ion etching (See 5. Method for the specific conditions). We note that the fabrication procedure used to realize AD-ZIM is much simpler than that we used to realize metal-clad ZIM [18].

Fig. 5(a) shows representative scanning electron microscope images of a fabricated device with the optimized parameters. 22 prism devices with different $2r$ were fabricated on the same chip, each measuring 8 pillars across. Light is coupled into the prism from a Si waveguide that is adiabatically widened to enable prism excitation with a nearly uniform wave-front. The light propagates through the prism with a phase velocity determined by the prism's effective index, and refracts at the output facet of the prism. The refracted light couples into a semicircular SU-8 slab waveguide, where it propagates to the edge and scatters out (See 5. Method). Therefore, by monitoring the intensity of the scattered signal at the output of the SU-8 waveguide, we can determine the refraction angle $\alpha$. Figure 5(b) shows the refracted light from one of the prisms, as measured by a near-infrared (NIR) camera positioned above the sample. Despite strong scattering from the prism itself, the refracted beam is clearly observed at the edge of the SU-8 slab waveguide at $\alpha = 0°$, consistent with our expectation of $n_{eff} = 0$ based on Snell's law.

Figure 6 summarizes the results for three samples with different $2r$. Figure 6(a) shows the angular intensity distribution of the scattered light taken along the edge of the SU-8 slab waveguide with different input wavelength from 1480 nm to 1680 nm. As we vary the input wavelength, the refraction angle $\alpha$ changes gradually and continuously (without a bandgap) from positive refraction angles at short wavelengths to negative refraction angles at long wavelengths. Between these two regimes, each sample clearly shows $\alpha = 0°$. The zero-crossing wavelength increases as $2r$ increases, in agreement with the theoretical prediction shown in Fig. 4. To estimate $n_{eff}$, we use Snell's law ($n_{eff} = n_{SU8} \cdot \sin(\alpha)/\sin(45°)$, where $n_{SU8} = 1.58$ is the index of SU-8 polymer). Because both our simulated and measured refraction beams have a spatial broadening due to diffraction, we perform 2D Gaussian fitting for the measured and simulated near field patterns [20]. Figure 6(b) compares the simulated and measured $n_{eff}$. Based on this comparison, we can conclude that we have successfully demonstrated the proposed AD-ZIM concept and its fabrication tolerance. Our experimental results are in the excellent agreement with the theoretical predictions. Small discrepancies between the two may be attributed to the material parameters used in the simulations that were measured before Si growth as well as due to the tapered profile of the etched pillars (inset of Fig. 5(a)) with an angle of $2 - 3°$.

## 4. CONCLUSION

We presented on-chip fabrication-tolerant all-dielectric zero-index materials (AD-ZIMs). With a proper choice of lattice constant $a$ and pillar height $h_{Si}$, the PDC achieves fabrication tolerance against variation of pillar diameter $2r$, which is obtained by engineering the structural dependencies of the target modes. We also demonstrated this concept by fabricating on-chip AD-ZIM prism devices with different $2r$, and measuring their $n_{eff}$. Fabrication simplicity and tolerance increases the yield ratio of ZIMs dramatically, enabling many applications of on-chip ZIM including unconventional phase-matching properties in nonlinear optics [6] with large area. This concept can be extended to other operation wavelength regimes with different materials by changing the overall scaling factor of the pillar array.

## 5. METHOD

For the finite element method simulation, we use commercially available software (COMSOL Multiphysics®) to find the eigenfrequencies of the unit cell as shown in the inset of Fig. 1(a). We apply periodic boundary conditions in $x$ and $y$ directions and 2-μm-thick PMLs in *z* direction. For the material parameters of the Si and SiO$_2$, we applied the material parameters measured using spectroscopic ellipsometry [18].

In terms of the fabrication, first we performed Si growth on a commercial SOI wafer with a 500- nm Si layer on a 3-μm SiO$_2$ layer. The native oxide was removed by brief BHF dip. Then, using a home-built chemical vapor deposition (CVD) chamber, we deposited Si with 5 mTorr SiH$_4$ partial pressure and temperature of 775 °C. The growth rate was ~ 30 – 40 nm/min. We have also checked the cross section after the growth, and defects between the original Si layer and the regrown Si layer were not observed by SEM. With this regrown SOI wafer, we perform three iterations of e-beam lithography (ELS-125, Elionix) to produce Si prisms/waveguides, bottom gold mirrors, and SU-8 waveguides. For the first iteration, we spin-coated HSQ negative resist (Fox16, Dow Corning) so that the resist thickness is around 300 – 400 nm, and patterned the AD-ZIM prisms and the coupling Si tapered waveguides. The resist pattern was transferred into the Si layer by using inductively coupled plasma reactive ion etching. The second lithography step was followed by a lift-off process with using poly (methyl methacrylate) (PMMA) positive resist (495PMMA A6, MicroChem) to form bottom gold mirrors for the refraction angle measurement. For the gold deposition, we deposited 5- nm-thick titanium and 45- nm-thick gold by using an e-beam evaporator. For the third lithography step, we spin-coated SU-8 polymer (SU-8 2002, MicroChem) and formed the SU-8 slab waveguides and SU-8/Si mode converters for the optical fiber input at the cleaved facet.

To carry out the measurements, we butt-coupled a lensed fiber to the cleaved SU-8 waveguide facets. We measure light scattering from the surface to calculate the refraction angle of AD-ZIM prisms using InGaAs NIR camera (SU640HSX-1.7RT, Goodrich) through x50 objective lens while sweeping the laser wavelength of telecom tunable lasers (TSL-510, Santec). To enhance the scattering light intensity at the edge of the SU-8 slab waveguide, we deposited semicircular strip of gold along the edge of SU-8 slab. We also patterned a reference SU-8 waveguide surrounding the SU-8 slab waveguide.

**Funding**

Air Force Office of Scientific Research (AFOSR) (FA9550-14-1-0389); National Science Foundation (NSF) (DMR-1360889); Natural Sciences and Engineering Research Council of Canada (NSERC), Harvard Quantum Optics Center (HQOC); Japan Society for the Promotion of Science (JSPS).

**Acknowledgment**

The authors thank S. Kalchmair, A. Shneidman, I. Huang, C. Wang, Y. Sohn, and Z. Lin for discussions. The authors thank Prof. C. Lieber for supporting Si regrowth process.

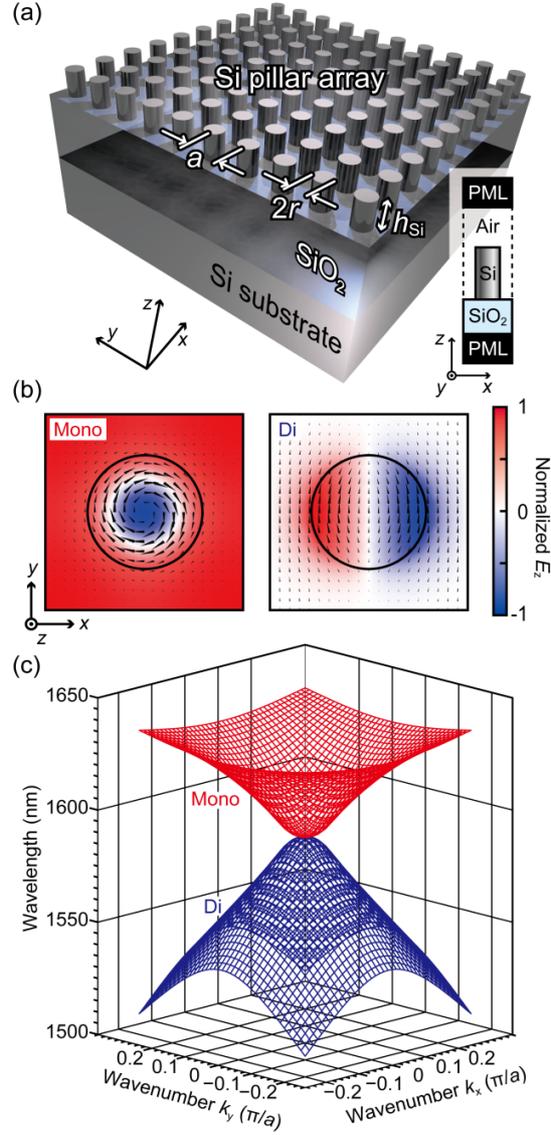

**Fig. 1.** On-chip fabrication-tolerant AD-ZIM in the telecom regime. (a) Schematic. The inset shows the unit cell sandwiched by perfectly matched layers (PMLs). (b) Representative $E_z$ (color) and $H$ (black arrows) distribution at the middle of the Si pillar in a single unit cell for electric monopole (left) and magnetic dipole mode (right) at the Γ point for wavelength of 1596.7 nm. $a$ = 918 nm, $h_{Si}$ = 860 nm, $2r$ = 528 nm were used. (c) Representative 3D dispersion surface of two modes with the $k$ vector ($k_x$, $k_y$) around Γ point ($a$ = 918 nm, $h_{Si}$ = 860 nm, $2r$ = 512 nm). Here, we omit the "dark" longitudinal magnetic dipole mode [1] (black dots in Fig. 4(b)) for clarity. ). The quality factors of monopole and dipole modes are $Q > 10^5$ and $Q \sim 50$, respectively. Different $Q$s result in different imaginary parts of modes' eigenfrequencies, even when there is degeneracy in the real parts, giving rise to the quadratic dispersion around Γ point. This is confirmed by a 2D model of our structure (not shown), with infinitely long pillars: lack of out-of-plane losses results in infinitely large $Q$s for both modes and consequently the linear dispersion around Γ point is recovered. We note that the effect of finite and different $Q$s is analogous to parity-time symmetry without gain [19]

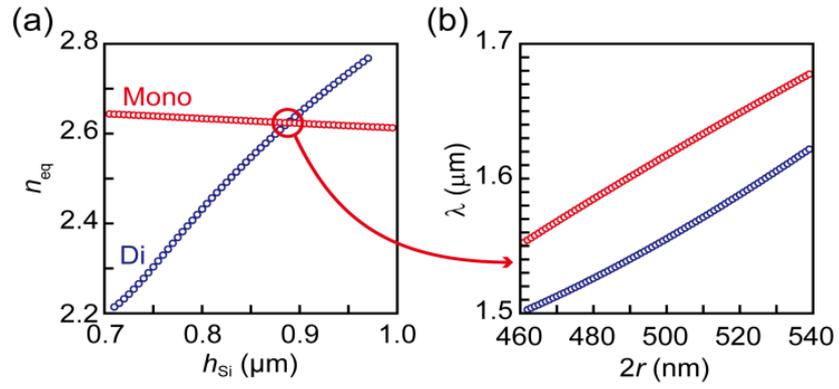

**Fig. 2.** (a) $n_{eq}$ for monopole (red dots) and dipole (blue dots) modes as a function of $h_{Si}$ ($a$ = 879 nm, $2r$ = 512 nm). The modes have identical $n_{eq}$ when $h_{Si}$ = 885 nm (circled point). (b) Wavelengths of two eigenmodes at $\Gamma$ point as a function of $2r$ for $h_{Si}$ = 885 nm. In this case, $\Delta\lambda/\Delta 2r$ of two modes are nearly identical for wide range of $2r$, as indicated by nearly parallel curves.

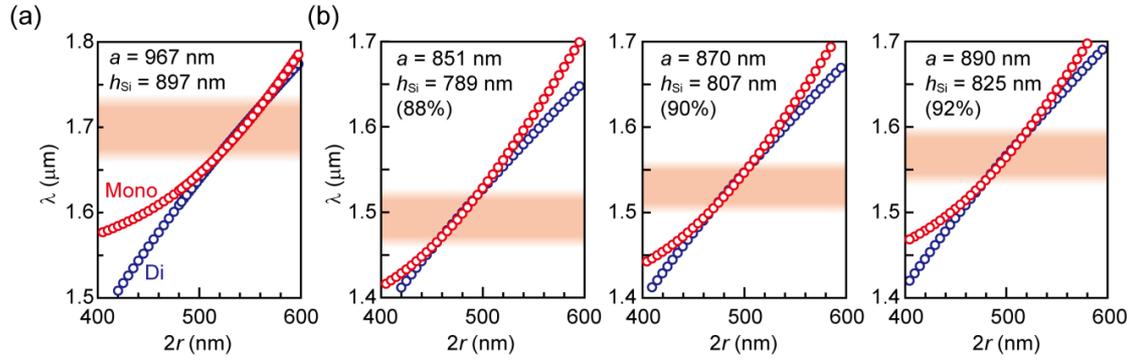

**Fig. 3.** Scaling laws for structural robustness in AD-ZIM. (a) $\lambda$ as a function of $2r$ with $a$ and $h_{Si}$ of monopole (red dots) and dipole (blue dots) modes obtained by solving Eq. 1. The orange area indicates the near-degeneracy regime ($\Delta\lambda < 1$ nm). (b) $\lambda$ as a function of $2r$ with all parameters scaled-down to 88%, 90%, and 92% of their values used in (a).

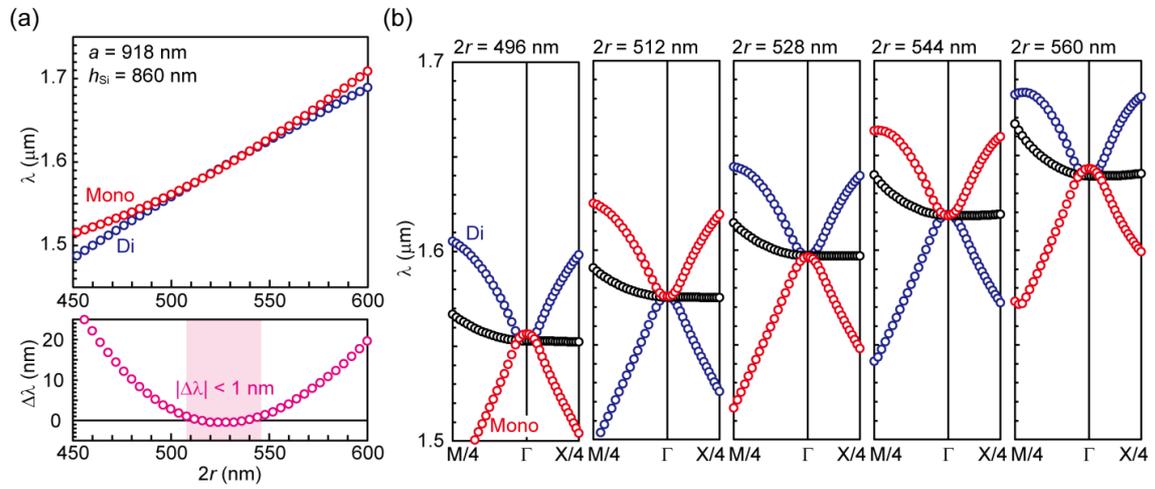

**Fig. 4.** Fabrication-tolerant PDC, insensitive to variations in $2r$. (a) $\Gamma$-point wavelengths for monopole (red dots) and dipole (blue dots) modes (top) and their difference $\Delta\lambda$ (bottom) with different pillar diameter $2r$. (b) Photonic band diagrams with five different $2r$ showing PDCs at different wavelengths. Black dots indicate "dark" longitudinal magnetic dipole mode [1].

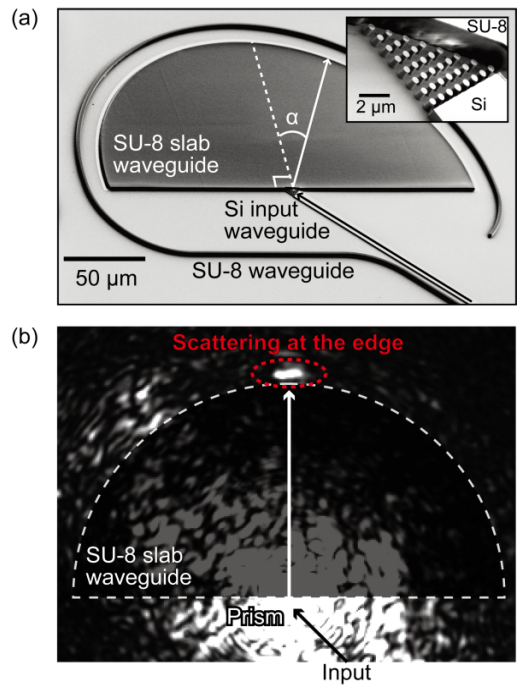

**Fig. 5.** Experimental demonstration of an on-chip fabrication-tolerant AD-ZIM prism integrated with Si photonics. (a) Scanning electron microscope image of the fabricated device. The inset shows the magnified picture of the prism. (b) Representative NIR image showing light scattered from the prism. The refracted beam with α ~ 0° is visible at the top. The white dashed line shows the outline of the SU-8 slab waveguide.

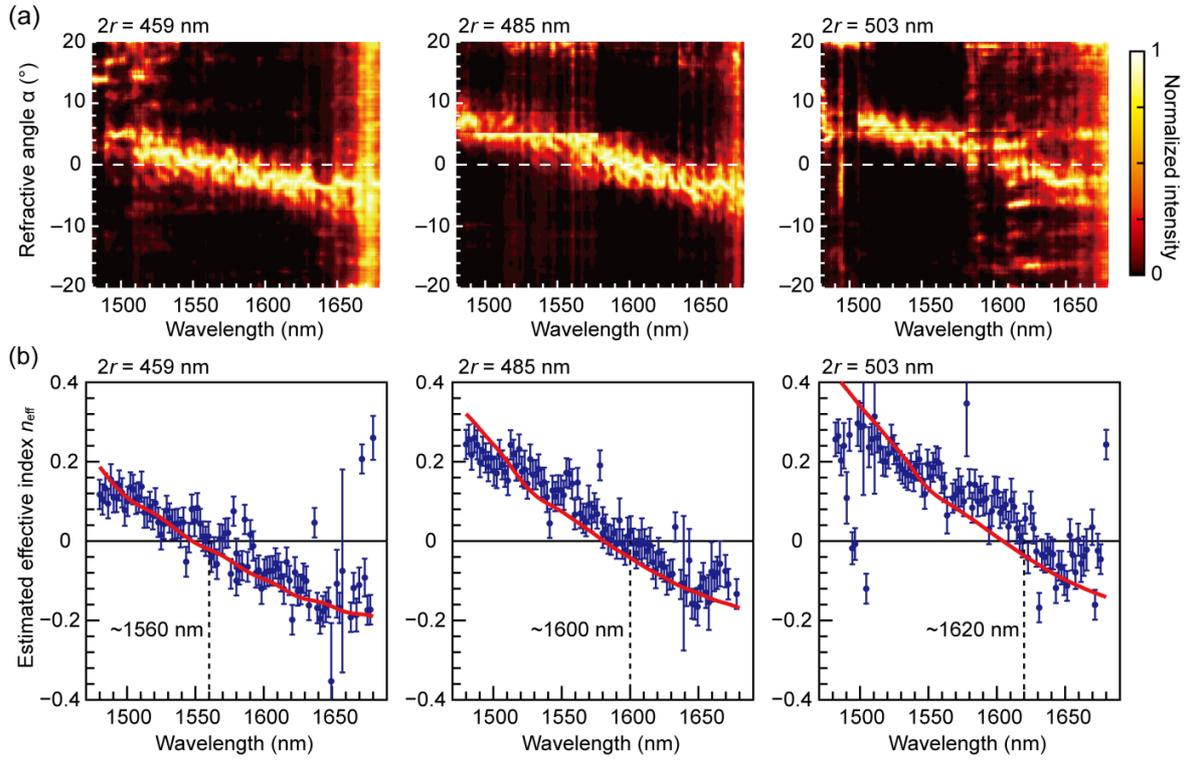

**Fig. 6.** Demonstration of fabrication-tolerant AD-ZIM ($a = 918$ nm, $h_{Si} = 860$ nm). (a) Observed angular intensity distribution of the scattered light along the edge of the SU-8 slab waveguide as a function of input wavelength for samples with different $2r$. (b) Estimated effective index $n_{eff}$ for (a). Blue dots with error bars and red lines indicate the experimental and theoretical results, respectively. The theoretical results are simulated using 3D finite difference time domain (FDTD) with the same geometries of the fabricated devices. Error bars depict uncertainties in the measurement.